# Multiscale fabrication of scalable biomimetic 3-D, integrated micro-nanochannels network in PDMS for solute exchange

Prasoon Kumar[123], Prasanna S Gandhi[2], Mainak Majumdar[3]

*1 IITB-Monash Research Academy, Powai, Mumbai, Maharashtra-400076, India*
*2 Suman Mashruwala Advanced Microengineering Laboratory, Department of Mechanical Engineering, Indian Institute of Technology Bombay, Powai, Mumbai, Maharashtra 400076, India*
*3 Nanoscale Science and Engineering Laboratory (NSEL), Department of Mechanical and Aerospace Engineering, Monash University, Clayton, Melbourne, Australia*

**Abstract**

Integrated micro-nanochannel networks in fluidic devices are desirable in a number of applications ranging from self-healing/cooling materials to bioengineering. The conventional micro-manufacturing techniques are capable of either producing microchannel or nanochannel networks for a fluidic application but lack proficiency in the production of integrated micro-nanochannel network with a smooth transition from micro-to-nano scale dimension. In addition, these techniques possess limitations such as heavy initial investment, sophistication in operation and scale-up capabilities. Therefore, the current paper demonstrates the combination of micro/nanotechnologies to design and develop a biomimetic 3-D integrated micro-nanochannel network in PDMS device for solute exchange. We have used 3-D printer, a scalable technology, to design and manufacture micro-mold having fractal-shaped features. Further, electrospinning was used to deposit nanofibrous network on the fractal mold. Subsequent micro-molding with PDMS was used to obtain fractal-shaped microchannels integrated with embedded nanofibers. Henceforth, solvent etching of nanofibers followed by bonding of thin PDMS membrane generated by spin coating to open end of channels lead to the formation of functional microdevices. These PDMS devices mimic the natural vasculature of living system, where fractal-shaped microchannels will assist in efficient fluid flow and the site of nanovascular network participates in heat/mass transport operations. Further, dye flow propounds the functionality of such devices. Our study hence proposes a simple and scalable hybrid microtechnolgy to fabricate fluidic devices having multiscale architecture. This will also facilitate the rapid fabrication of microfluidic devices for biomedical, diagnostics, sensors and micro-TAS applications.

## 1. Introduction

Integrated three dimensional micro-nanochannel networks in polymer matrix have become significant owing to their potential applications in filtration, self-healing and cooling materials, tissue-engineering and regenerative medicine [1]. However, technologies are available to fabricate such polymer system with either nanoscale or microscale dimensions [2]. With the advent of additive micro manufacturing techniques, it has become possible to generate microstructures with resolution up to 1μm [3-4]. Moreover, challenges associated with successful integration of different micro/nano manufacturing techniques are their different operating conditions, compatibility with substrate materials and capabilities to form structures with smooth transition from meso-to-micro-to-nano domain. Further, even if such criteria are met under certain conditions, it does not compile a method to produce integrated micro-nanostructures in polymeric substrate on a large scale [5]. In addition, these fabrication techniques require huge investment for the installation and supporting infrastructure. Further, being highly sophisticated techniques; they require high technical expertise, time and resources; therefore their tailor ability to suit any hybrid micro manufacturing is quite challenging [2]. Although other recently explored techniques like direct self-assembly of fugitive inks, electrical discharge, have demonstrated the formation of vascularization in polymer matrixes with much ease, their control over the dimensions of vascular channels and scalability remain an intrinsic limitation [6-8]. Therefore, recent researchers have demonstrated the application of two scalable additive technologies; 3-D printing and electrospinning to generate integrated micro-nanostructure to be used as a scaffold for tissue regeneration [9-10]. However, no such attempt has been made to the best of our knowledge to integrate 3-D printing and electrospinning to generate micro/nanofluidic devices for lab-on-a-chip or organ-on-a-chip applications. Hence, in the current study, we have integrated scalable micro-manufacturing technologies: electrospinning, 3-D printing, micro-molding and solvent etching of sacrificial structures in an innovative way to design and develop a PDMS based micro-devices having integrated micro-nanochannel networks and a thin membrane for convection-diffusion studies . These devices mimics natural vasculature of living beings where fractal-shaped microchannel network assist in efficient fluid flow, the site corresponding to nanovascular network participates in heat/mass transport operations and thin membrane assists in heat/mass transfer operations.

## 2. Materials and method

### 2.1. Materials

Polydimethysiloxane (PDMS) (Sylgard® 184 silicone *elastomer* kit - *Dow* Corning) was used for fabrication of micro/nanofluidic devices. Polystyrene (PS) 192 KDa ( Sigma Aldrich Pvt. Ltd., India) was used for the fabrication of electrospun micro/nanofibrous mesh. N, N, Dimethylformamide (DMF) ( Ajayx PVT. Ltd, Australia) was used as solvent for PS solution and sacrificial etching of micro/nanofibrous structures, Polycarbonate (PC) (Stratasys, Eden Prairie, MN, USA ) was used to fabricate 3-D molds with 3-D printer, Fluorecin Dye (Sigma Aldrich Pvt. Ltd., India) was used as dye in dye flow experiments. Ethanol (Changsu Yangyuan Chemicals, China) was the fluid flown from fractal-shaped microchannels to micro/nanochannel networks during dye flow experiments.

### 2.2. Methods

Electrospinning (designed and assembled in lab) was used to fabricate polystyrene nanofibrous mesh [11]. For electrospinning, a 15% w/v of polystyrene solution was prepared in DMF. The solution was stirred overnight to form a homogenous solution before electrospinning. The process parameters were optimized to generate fibres with 1μm diameter: voltage – 10KV, distance between collector and spinneret – 12cm, flow rate of solution by syringe pump – 0.5ml/hr, deposition time 1–2 minutes, and needle gauge of 24G. The micro/nanofibrous mesh of polystyrene were obtained on aluminum foil. A part of the sample were taken for imaging and characterization by scanning electron microscope (SEM) while other part were used in further fabrication process. In parallel, a mechanical design of the device mold with fractal-shaped microstructures were developed in Pro/ENGINEER Wildfire 4.2 and subsequently, used as an input to 3-D printer (Objet. Eden260V™, Stratasys, Eden Prairie, MN, USA) to generate a structure in polycarbonate (Figure 2). The mold was carefully cleaned with ethanol and water in a water bath sonicator. Further, electrospun nanofibrous mesh deposited on aluminum foil was carefully transferred to the mold.

In order to avoid displacement of fibrous mesh during micro-molding process by PDMS, few drops of ethanol was added over the fibrous mesh placed in a mold to enable stiction of fibrous mesh upon evaporation to the mold structure (Figure 1). Thereafter, PDMS (10:1) was poured into the mold and cured over a hot plate at 70$^o$C for 6-8 hours. After cooling of the mold, PDMS devices having embedded nanofibrous mesh were placed in a DMF bath and stirred slowly to enable etching of fibrous structures for several days. After an etching process, an inverse replica of porous non-woven electrospun mesh integrated with fractal-shaped microchannels were created in a cured PDMS block of 5mm thickness.

Further, PDMS solution with base and curing agent in 10:1 was prepared and spin coated over paraffin wax paper with a spin coater (Spin 150-v3, SPS, Europe, Inc) with rpm varying from 1000 rpm to 5000 rpm. The coated films were cured at room temperature (25$^o$C-30$^o$C) for 48 hrs. The thickness of PDMS membrane was estimated by a screw-guage. Thereafter, open end of the PDMS device having integrated micro-nanochannel networks was plasma bonded to PDMS membrane generated previously on paraffin wax paper in a plasma cleaner (Basic Plasma Cleaner PDC-32G, Harrick Plasma, USA) at 1mbar for 20second. The device was heated at 70$^o$C on hot plate for 20-30 minutes to further strengthen the plasma bonding. Thereafter, few samples were taken for SEM and optical imaging to characterize the dimensions of different channel features in the device. The dimensions of the different channel features were estimated by image processing in MATLAB 7.4. Further, holes were punched, tubing were connected to the above PDMS device to be used for fluid flow by capillary action.

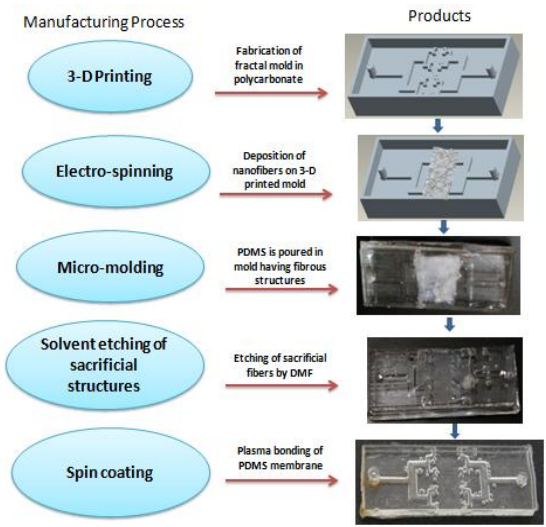

**Figure 1. Schematic of process of device manufacturing**

## 3. Results and discussion

The computer aided design (CAD) model of fractal-shaped microstructure were created till second generation branches. The model assumes the width of the fractal structures from parent to 2nd generation varies from 1mm to 0.447mm following the $W = W_o N^{-0.733}$ while length varies from 5mm to 0.921mm following the $L = L_o N^{-1.54}$. However, when the final mold was fabricated with 3-D printer in polycarbonate, it was observed that the dimensions of the structures were nearly same for parent and 1st generation fingers (Figure 2) while the morphology of 2nd generation fingers were tampered. This might be due to resolution limitation of the 3-D printer. Further, the height of the fractal structure came around 270 µm (Figure 2) after fabrication while CAD model input was given height of 250 µm.

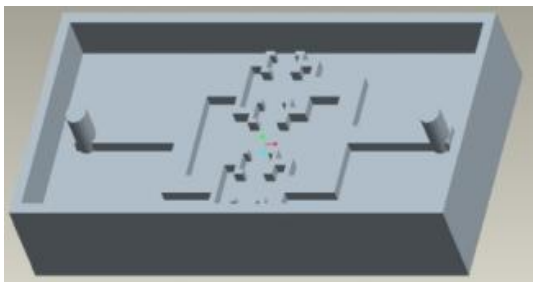

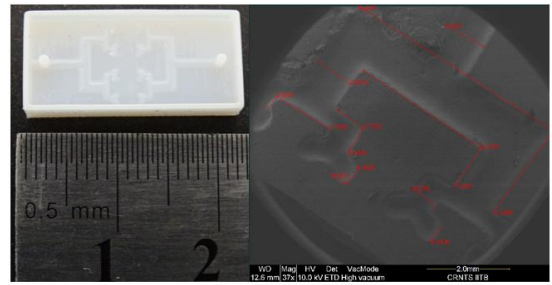

**Fig. 2. ProE design of mold (Top) and 3-D printed mold in polycarbonate (bottom left) and SEM image of mold with feature dimensions in millimeter (bottom right)**

Further, the SEM of nanofibrous meshes revealed the dimension of fibers to be 1.304±0.318µm (Figure 3A) while SEM of PDMS devices showed successful solvent etching of polystrene nanofibrous mesh to generate nanochannel network (Figure 3B). The dimensions of channel network were in corroboration with the dimensions of micro-nanostructures used in creating inverse replica in PDMS. This suggested the non-swelling activity of DMF on PDMS. The PDMS device having open ended fractal-shaped microchannel with open ended integrated nanochannels were covered by PDMS membranes. The SEM of cross-section of PDMS device bonded with thin membrane were taken at two different positions. The thickness of membrane bonded to device come around 38µm and figure 3C suggest the continuous integration of membrane with PDMS by plasma bonding. Further, SEM of cross-sectioned PDMS device at nanochannel networks suggested that nanochannel layer spanned around 166µm (Figure 3D). However, the thickness of membrane estimated came around 51µm. This might be because of difficulty in demarcating the point of joining of PDMS membrane and PDMS device.

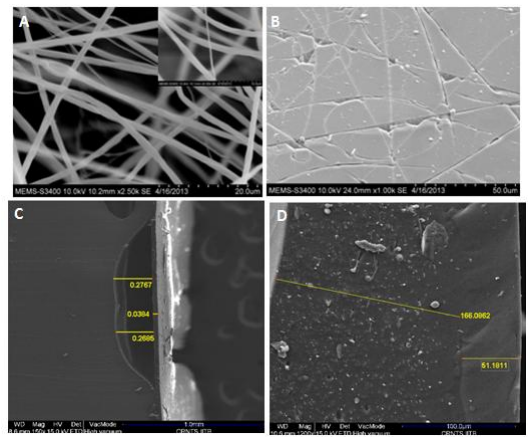

**Figure 3 SEM image A) electrospun nanofibrous mesh B) nanochannel network in PDMS C) cross-section of fractal channel with membrane D)**

**cross-section of nanochannel network with membrane**

These membranes were spin coated with PDMS solution on a paraffin wax paper and cured at room temperature before being plasma bonded with PDMS devices (Figure 3C and 6). The presence of paraffin wax enabled easy transfer of membrane to PDMS device. The graph in figure 4 suggest the further lowering of membrane thickness by increasing the speed of rotation during spin coating. However, very thin membranes may be difficult to transfer from paraffin wax to PDMS device. This process enabled fabrication of membranes with a thickness as low as 15μm. Further, spin coating over paraffin film may assist in the generation of large sized thin membranes and easy integration with PDMS devices during scale up device fabrication.

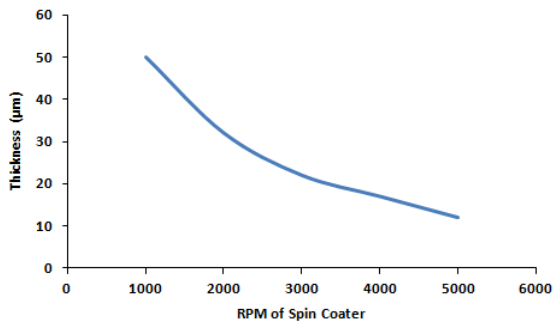

**Figure 4 Graph showing the variation of thickness of membrane with speed of rotation of spin coater**

Further, Fluorecin dye in ethanol was used as fluid and flow experiments were carried out by capillary action. It was observed that the dye could easily flow in the fractal fingers as shown in figure 5 but we could not observe the dye solution in nanochannel networks. Thus, further experimentation with dye flow through these devices and solute exchange by convection diffusion mode will establish the design and fabrication of micro/nanofluidic devices by method proposed in paper for various applications.

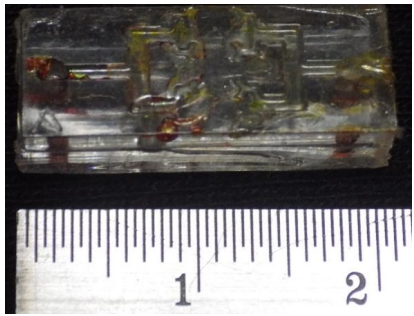

**Figure 5 Image of Dye presence in channels of PDMS device**

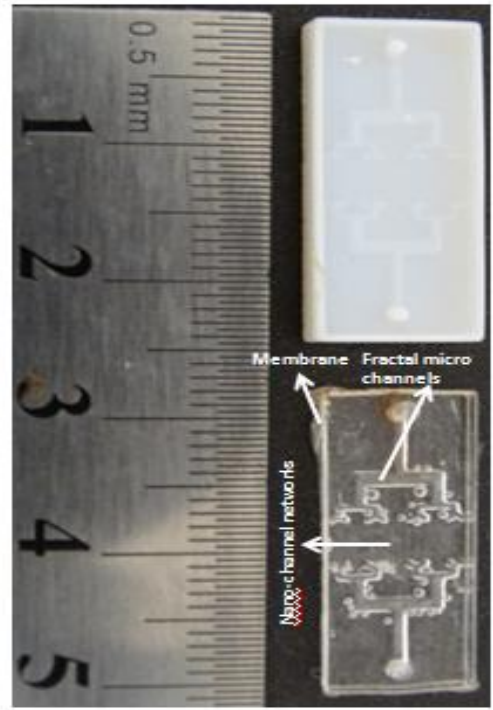

**Figure 6 Image of PDMS device with fractal-shaped microchannels integrated with nanochannel network**

## 4. Conclusions

The hybrid additive micromanufacturing techniques are capable of producing integrated micro-nano structures with wide span in their dimensional variability. Therefore, stand-alone technologies are currently being replaced by a combination of microtechnologies for complex applications. We have demonstrated the combination of additive micro/nanotechnologies to design and develop a biomimetic 3-D integrated micro-nanochannel network in PDMS device for solute exchange. We have used 3-D printer and electrospinning, scalable technologies, to design and manufacture micro-mold having fractal-shaped features and nanoscale structures respectively in a device. Our unique way to combine the above scalable technologies with micro-molding and sacrificial etching lead to generation of 3-D integrated micro-nanochannel network in PDMS device, good substitute for conventional lithography techniques. We are currently working towards the fluid flow and solute exchange through such devices. These PDMS devices are a natural mimic of vasculature system of living beings, where fractal-shaped microchannels enable efficient fluid flow whereas nanovascular network serves as the site for heat/mass transport operations. Thus, our study suggests a simple and scalable hybrid microtechnolgy to fabricate fluidic devices having multiscale architecture. This will also facilitate the rapid fabrication of microfluidic devices for biomedical, self-healing/cooling materials, filteration, diagnostics, sensors, and micro-TAS applications.

**Acknowledgements**

Authors would like to acknowledge the IITB-Monash Academy for financial support. Authors would also like to thank SAIF (IIT Bombay) for SEM characterization and Materials Science and Engineering Department for use of 3-D printer